\documentclass[conference]{IEEEtran}
\IEEEoverridecommandlockouts
\usepackage{cite}

\usepackage{hyperref}   
\usepackage{subcaption}
\usepackage{multirow}
\usepackage{booktabs} 
\usepackage{easyReview}
\usepackage{amsmath,amssymb,amsfonts}
\usepackage{algorithmic}
\usepackage{graphicx}
\usepackage{textcomp}
\usepackage{xcolor}

\def\BibTeX{{\rm B\kern-.05em{\sc i\kern-.025em b}\kern-.08em
    T\kern-.1667em\lower.7ex\hbox{E}\kern-.125emX}}
\begin{document}

\title{Effectiveness of High-Dimensional Distance Metrics on Solar Flare Time Series\\
}

\author{
    \IEEEauthorblockN{Elaina~Rohlfing\IEEEauthorrefmark{1},
    Azim~Ahmadzadeh\IEEEauthorrefmark{1},
    V.~Aparna\IEEEauthorrefmark{2}\IEEEauthorrefmark{3}}
    \IEEEauthorblockA{
        \IEEEauthorrefmark{1}Department of Computer Science,
        University of Missouri-St. Louis,
        USA\\
        \IEEEauthorrefmark{2}
        Bay Area Environmental Research Institute, NASA Research Park, Moffett Field, CA 94035, USA \\
        \IEEEauthorrefmark{3}
        Lockheed Martin Solar and Astrophysics Laboratory, 3251 Hanover Street, Building 203, Palo Alto, CA 94306, USA \\
        Email: \IEEEauthorrefmark{1}emrwc6@umsl.eduu \\
        \vspace{-1.1cm}
    }
}

    
\maketitle

\begin{abstract}
Solar-flare forecasting has been extensively researched yet remains an open problem. In this paper, we investigate the contributions of elastic distance measures for detecting patterns in the solar-flare dataset, SWAN-SF. We employ a simple $k$-medoids clustering algorithm to evaluate the effectiveness of advanced, high-dimensional distance metrics. Our results show that, despite thorough optimization, none of the elastic distances outperform Euclidean distance by a significant margin. We demonstrate that, although elastic measures have shown promise for univariate time series, when applied to the multivariate time series of SWAN-SF, characterized by the high stochasticity of solar activity, they effectively collapse to Euclidean distance. We conduct thousands of experiments and present both quantitative and qualitative evidence supporting this finding.

\end{abstract}

\begin{IEEEkeywords}
time series, flare, distance, clustering
\end{IEEEkeywords}
%
%
%
%
%
%
%
%
%
%
%
%
%
%
%
%
%
%
%
%
\section{Introduction}\label{sec:intro}

    This paper focuses specifically on the Space Weather Analytics for Solar Flares (SWAN-SF) dataset, a benchmark for developing solar flare forecasting models. Solar flares are intense bursts of radiation from stored magnetic energy, occurring on the Sun, that can release energy to the order of $10^{32}$ ergs, equivalent to 2000-million Megatons of TNT (the atomic bomb over Nagasaki produced energy equivalent to that from 20 kilotons of TNT). Flares are classified as A, B, C, M and X-class flares with the energy released increasing by ten times with each letter starting from $10^{28}$ ergs for A-class to $10^{32}$ ergs of energy for X-class flares \cite{2009ApJ...700..199T,2005JGRA..11011103E}. Each type of flare is further divided by numbers 1 through 9. The intense radiation in extreme-ultraviolet (EUV) and X-ray from stronger flares (M and X-class) can cause heating and expansion of the Earth’s atmosphere that can affect attitudes of low-Earth satellites by increasing their drag. Additionally, the variable charging of the ionosphere during flaring events can disrupt radio communication and navigation signals due to disruption in the ionospheric propagation pathways \cite{2023Astro...2..165M}. For example, recently during February 2022, several communication satellites launched by Space Exploration Technologies Corporation (commonly known as SpaceX) plummeted back to Earth due to increased atmospheric drag \cite{2023Astro...2..165M,2022SpWea..2003074H}. Stronger flares, especially X-class ones, are almost always associated with a coronal mass ejection \cite{2005JGRA..11012S05Y}. CMEs are drivers of geomagnetic storms by inducing disturbances in the Earth’s magnetosphere that, in turn induce currents in metal pipelines on and underground including the electric grid systems. Examples include the famous Hydro-Quebec blackout event due to the March 1989 geomagnetic storm induced by two back-to-back CMEs each associated with an X4.5 and an M7.3 flare \cite{2019SpWea..17.1427B}. More recently, the Gannon storm of May 2024 that severely affected the navigation satellites and caused millions of dollars losses to agriculturalists \cite{koebler2024} were caused by several CMEs associated with X-class flares as well \cite{2024SpWea..2204126T}. If a Carrington type event (occurred in 1859 and caused telegraph wires to catch fire due to induced currents; Carrington, 1859), were to occur today, it is expected to incur an unrecoverable economic loss of up to 42 billion dollars for the US alone \cite{2017SpWea..15...65O}.
    
    The forecasting of these events is thus a high-stakes problem where errors have significant consequences. The ideal forecasting method would minimize both false alarms and misses, but that is not an easy task. Depending on the usage, it may be acceptable to prioritize only one. A ``miss,'' where a flare occurs without a prior warning, exposes astronauts to dangerous radiation. Conversely, a ``false alarm,'' where a flare is predicted but does not materialize, can lead to costly operational disruptions, such as halting extravehicular activities on space missions. Despite numerous research, reliably predicting these events remains an open challenge.

    The SWAN-SF dataset has served as a test bed for flare forecasting algorithms during the past few years. A quick perusal of studies using SWAN-SF confirms the continuation of forecast stagnation in this domain \cite{hostetter2019understanding, ahmadzadeh2019challenges, ahmadzadeh2019rare}. This is largely due to the stochasticity of the pre-flare activities, but also, due to the complexity of high-dimensional data. Note that, SWAN-SF is a multivariate time series dataset and considering only its 24 physical parameters, each observation window is described by 1,440 quantities. This so-called ``curse of dimensionality'' was recognized decades ago \cite{koppen2000curse, verleysen2005curse} recognizing $d>6$ as high dimensional---we deal with over one thousand dimensions! For this reason, a significant portion of existing research bypasses the temporal dynamics of flares entirely, replacing the time series with their summary statistics (e.g., median, standard deviation, etc.) \cite{wen2022improving, ahmadzadeh2021how} rather than engaging with the raw series data \cite{ma2019solar, ma2019segmented, ma2017distance}. This transformation loses information, but the hope is that it preserves the valuable information and what is lost is mostly noise \cite{hostetter2019understanding}.
    
    With this introduction, we raise a question: are advanced high-dimensional distance measures effective for flare forecasting? Specifically, do high-dimensional distance measures (which show promising results in other domains) capture similarities beyond the simple Euclidean distance? This is an important question because many machine-learning algorithms rely on similarity patterns to classify data points. Also, this question is not trivial because the models and metrics used should be optimized fully, leaving no room for suppressing the effectiveness of those metrics.

    To investigate this question, we use clustering algorithms, specifically the $k$-medoids algorithm. The choice to apply a clustering algorithm to a labeled dataset is a deliberate one. Clustering algorithms directly use distance measures to group time series based on their morphological similarity, and form clusters accordingly. Further, when a data point is labeled correctly or incorrectly, its distance from the medoids gives direct justification. By visualizing how such a distance was measured (e.g., seeing the point matching in Dynamic Time Warping measure), the effectiveness of the distance measures can be investigated. This feeds directly into the objective of this paper. Therefore, while we fully optimize our flare-forecast model, we deviate our focus from achieving higher performance quantities and focus on the effectiveness of the utilized metrics. 
    
    We focus on $k$-medoids algorithm (among many other clustering algorithms) following the recommendation of a recent study in which the authors demonstrate $k$-medoids to be especially well-suited to time series clustering tasks with elastic distance measures \cite{holder2023clustering}.

\section{Background}
    \subsection{Current State of Flare Forecasting}
        In the past decades, several data-driven algorithms have been used to advance our flare-forecast capabilities. Two of the most recent survey studies review the advancements and the challenges \cite{Georgoulis2024prediction, whitman2023review}. To provide context for this paper, we focus on the performance of flare forecast algorithms. Although in some studies, the reported performances are very high, others fail to reproduce such efforts---indicating flaws in some quantitative methods. For example, the models examined carefully in \cite{Leka2019comparison} never enjoy simultaneous TSS and HSS (or ApSS; see Sec.~\ref{subsec:evaluation} for their definition) of greater than 0.5, while in several other studies the inflated TSS of greater than 0.9 has been reported. Even when advanced ML algorithms are used, for $\geq$M-class flares, the subtle and realistic performance of TSS=0.46 and HSS=0.33 was reported \cite{vysakh2023solar}. The inflated cases may be rooted either in the use of small datasets which do not represent the population, or in the existence of some sort of information leakage during training and validation of models. A few of such flawed practices are reviewed in \cite{ahmadzadeh2021how}. Among the well-cited studies, \cite{bobra2015solarflare} used Support Vector Machines to predict flares and achieved TSS=0.61 and HSS=0.63 (for 24 hrs prediction). Although these numbers are realistic, the flaring class was defined as $\geq$ B (containing X-, M-, and B-class flares), whereas in other studies which report lower (overall) scores, the flaring class is defined as $\geq$M. The latter definition creates a more challenging problem and the studies often report lower scores for such a definition. All in all, it is evident that, under realistic assumptions, as TSS exceeds 0.5 (roughly speaking) HSS drops. Therefore, combinations similar to TSS=0.8 and HSS=0.2 are very common, and easy to achieve.

    \subsection{Time Series Clustering}

        Time series clustering is an unsupervised process which groups time series according to some notion of similarity. There are several classes of clustering algorithms. Clusters may be formed by finding hierarchical relationships, based on density, or by creating partitions of data points grouped around an ideal center. This paper utilizes the partitional clustering algorithm $k$-medoids. $k$-medoids is similar to the popular k-means. Both methods require $k$ to be specified and both allow different distance measures to used for finding grouping that minimizes distances from $k$ centers. This flexibility allows us to compare multiple distance measures with a single clustering method. While both algorithms form clusters by grouping data around a central point, their centers are defined differently. K-means chooses centroids that represent the exact center of its clusters. A centroid represents the true center of a cluster, but is usually only an approximation of the points in a dataset. $k$-medoids chooses medoids and requires them to be actual data points. Medoids are often the best approximation of the cluster center, rather than the exact center, but a medoid is always a representative from the dataset. Whether this is an advantage often depends on the characteristics of the data set in question. The study mentioned in Sec.~\ref{sec:intro}  \cite{holder2023clustering} demonstrates, in the case of univariate time series and elastic distance measures (see Sec.~\ref{sec:distances}), $k$-medoids strategy does offer a statistically significant advantage over k-means. The writers argue that $k$-medoids strategy preserves the nuanced temporal alignments that elastic distance measures are designed to capture, while the centroids produced by k-means lose those important details.

    \subsection{High-Dimensional Distance Measures}\label{sec:distances} 
    
        \begin{figure*}
            \begin{subfigure}[h]{0.32\linewidth}
                \includegraphics[width=\linewidth]{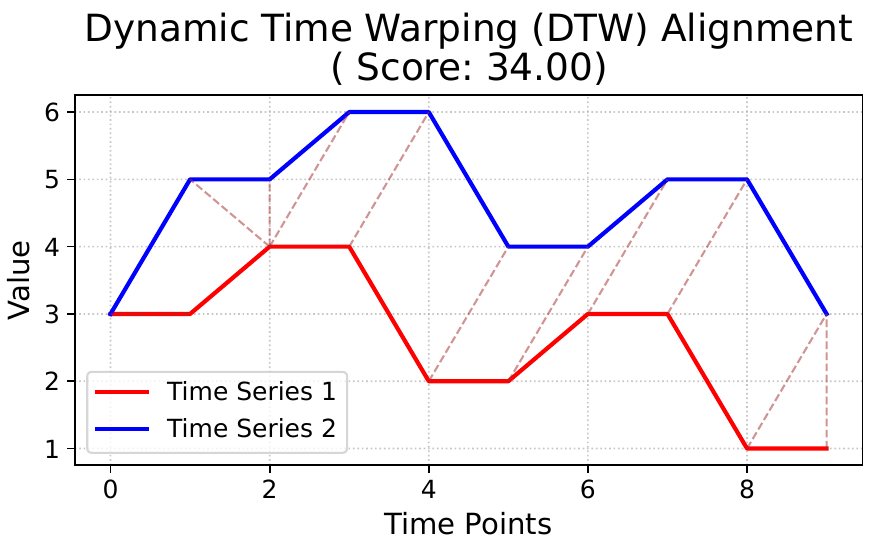}
            \end{subfigure}
            \hfill
            \begin{subfigure}[h]{0.32\linewidth}
                \includegraphics[width=\linewidth]{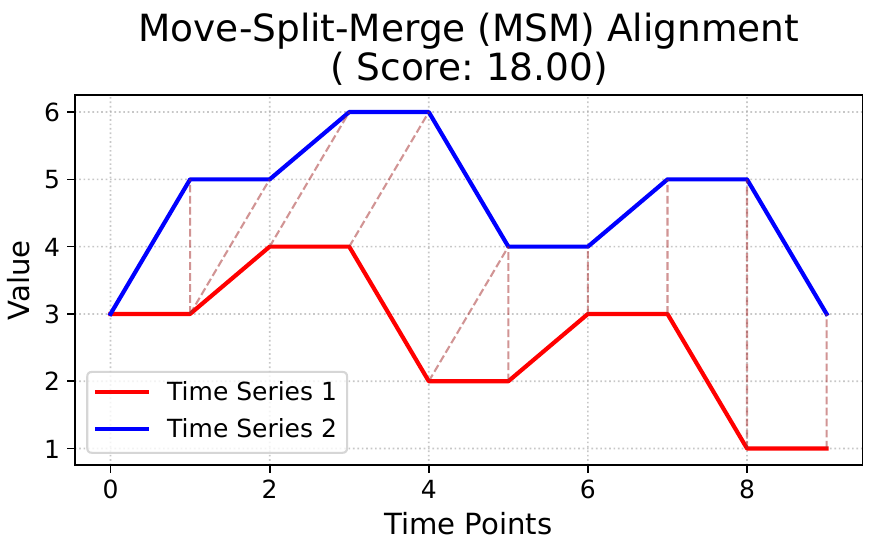}
            \end{subfigure}
            \hfill
            \begin{subfigure}[h]{0.32\linewidth}
                \includegraphics[width=\linewidth]{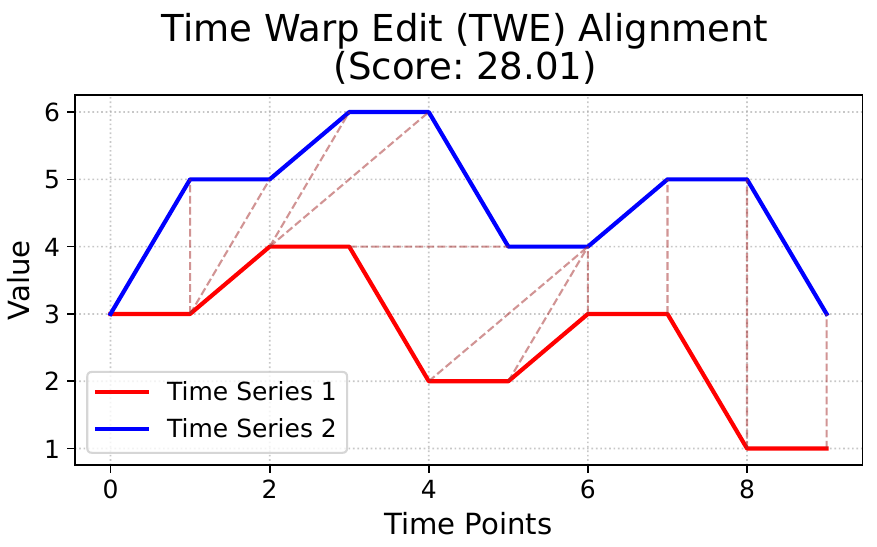}
            \end{subfigure}
            \caption{\footnotesize Illustration of alignment (point mapping) of DTW (left), MSM (middle), and TWE (right), when path is constricted by Sakoe-Chiba band ($w_{sc}=0.2$).}
            \label{fig:mapping_example}
        \end{figure*}
        
        In this paper we consider three elastic distance measures, Dynamic Time Warping (DTW) \cite{Berndt1994using}, Move-Split-Merge (MSM) \cite{stefan2013msm}, and Time Warp Edit (TWE) \cite{marteau2009twe} as well as the simple Euclidean Distance (EuD). A recent comparative analysis found these MSM and TWE to be the top performers out of nine elastic distance measures evaluated for time series clustering via $k$-medoids \cite{holder2024review}. MSM in particular, consistently outperformed the other eight measures, including the well-established DTW and several of its variants (Derivative DTW, Weighted DTW, and Weighted Derivative DTW). Due to their lower performance we exclude these variants of DTW from our analysis. Below, we review the measures.
        
        DTW is the most well known and thoroughly researched elastic distance measure for time series classification \cite{Berndt1994using, Ratanamahatana2005three}. For these reasons, DTW is a common benchmark in classification studies. DTW utilizes a warping technique to find the optimal alignment between the points of two series. Similarity is measured by computing EuD between these aligned points. 
        It has a single parameter, $w$. This parameter controls the width of the warping window, enforcing a boundary that constrains the warping path. Considering the cost matrix, $w$ restricts how far from the diagonal a warping path may move. A value between 0 and 1 determines the radius of the window size relative to the time series length. The most common types of warping windows are the Sakoe-Chiba band ($w_{sc}$) and Itakura Parallelogram ($w_{ip}$). $w_{sc}$ restricts the warping path to a uniform band along the diagonal, while $w_{ip}$ allows more warping near the center of the time series, but is more restrictive near the end points. When $w=1$, $w$ is global ($w_g$), meaning that warping is not constrained. Though $w$ is optional, in practice a constraint is almost always enforced. Unconstrained warping increases the likelihood of pathological warping, which is rarely desirable. Constraining DTW has repeatedly demonstrated a significant improvement in accuracy. When $w=0$, no warping is allowed and thus, DTW is reduced to EuD.
                
        MSM is a type of edit-distance measure, introduced in \cite{stefan2013msm}. 
        Like DTW, MSM measures similarity in terms of aligned points, but unlike DTW, MSM aligns points via edit operations. In this case, similarity is the minimized cost of transforming one series into another. Typical edit distance operations include $substitue$ or $match$, $insert$, and $delete$ and the cost of these transformations depends on the implementation of the specific edit distance. MSM redefines these basic edit operations so that the transformation cost can be made relative to the points affected. MSM's  operations are $move$, analogous to $substitue$, where a point replaces one of its neighboring values, $split$, similar to $insert$, but a point is split into two identical points, and $merge$, similar to $delete$, but a point is merged with an identical neighbor. Costs for these transformations are weighted according to how similar the point is to its neighbors. The idea is that a transformation involving a neighbor with a much higher or lower value should be penalized more than a transformation involving a neighbor with a similar or equal value. MSM has a single parameter $c$ which multiplicatively controls the severity of the transformation costs. Since $c$ is a constant multiplicative penalty, the default setting $c=1.0$ is neutral, equivalent to no additional penalty.

        TWE distance is another type of edit-distance measure \cite{marteau2009twe}. TWE is an edit distance that includes a warping component. The objective is to find the lowest cost transformation and warping path between two series. Like MSM, TWE provides its own implementations for the basic edit operations. TWE's edit operations are $match$, which transforms one segment of a series into another and $delete\_a$ and $delete\_b$, which eliminate a point from either series $a$ or series $b$, respectively. The cost associated with each edit operation is relative to the size of the vector(s) affected. TWE has two parameters, $\nu$ to control the stiffness of warping and $\lambda$, a constant penalty applied to edit operations. Although $\nu$ controls warping behavior, it does not bound the warping path (like DTW's $w$ does). Instead, it adds a multiplicative penalty to discourage extreme warping, increasing the cost of $match$ operations as the path moves further from the diagonal. TWE with no stiffness ($\nu=0$) is comparable to DTW, while TWE with infinite stiffness ($\nu=\infty$) is comparable to Euclidean distance. $\lambda$ controls the severity of transformation costs by adding a constant multiplicative penalty and is equivalent to MSM's $c$ parameter.

\section{Benchmark Dataset \& Evaluation Methods}\label{sec:data_and_evaluation_methods}
    In our experiments, we use a benchmark dataset to ensure the reliability and reproducibility of our findings. Further, we report the performance of our models using the metrics widely used by the community for comparability of our results. In the following sections, we present the dataset and the metrics.

    \subsection{SWAN-SF: A Flare-Forecast Benchmark Dataset}
        The Space-Weather Analytics for Solar Flare (SWAN-SF) dataset, is a multivariate time series dataset used to help researchers develop and evaluate tools to predict solar flare events \cite{angryk2020multivariate} (publicly available online \cite{angryk2020multivariate-data}); it serves as a test bed for flare-forecasting algorithms, making a relatively fair comparison of performance possible. SWAN-SF contains over four thousand multivariate time series, monitoring 51 flare-predictive parameters, between 2010 and 2018. This dataset is split into five partitions (P1-P5), with the objective of having roughly the same number of X- and M-class flares in each partition. The class-imbalance ratio varies between partitions as follows: 1:58 (P1), 1:69 (P2), 1:20 (P3), 1:51 (P4), and 1:95 (P5). Each multivariate time series instance is labeled as one of the five different classes, namely, X, M, C, B, and FQ. For dichotomous tasks, the X- and M-class flares represent the flaring (FL) class and the remaining classes represent the non-flaring (NF) class.
        
        There are a few studies which have narrowed down the 51 flare-predictive parameters of SWAN-SF to a more manageable size. In this study, we use ten of the most significant parameters as ranked in \cite{bobra2015solarflare, yeolekar2021feature}. Specifically, we use TOTUSJH, TOTBSQ, TOTPOT, TOTUSJZ, ABSNJZH, SAVNCPP, USFLUX, TOTFZ, MEANPOT, and EPSZ. For the exact definitions and formulas please see \cite{angryk2020multivariate}.

        Since the time series in SWAN-SF are collected using the rolling-window method, the overlap between consecutive instances may cause information leakage when randomly sampling to create the training, validation, and test sets. This phenomenon was addressed in \cite{ahmadzadeh2021how} and referred to as \textit{temporal coherence}. To prevent from information leakage due to temporal coherence in the data, P1 is selected for training, P2 for validation (tuning phase), and P3 is used for testing.
        
        As mentioned earlier, SWAN-SF intrinsically exhibit an extreme class-imbalance issue. We treat this by undersampling our training partition. Our undersampling method balances the FL (X, M) and NF (C, B, FQ) classes, while preserving the climatology of the flare classes. This \textit{climatology-preserving sampling} strategy (recommended in \cite{ahmadzadeh2021how}), produces a training set containing all FL-class instances (165 X and 1,129 M) and a subset of all NF instances, i.e., 1,294 NF instances (102 B, 115 C, 1077 N). This subset is $1.79\%$ of all NF instances in this partition. We apply no undersampling on the validation (P2) and test (P3) partitions.
        
        Lastly, the time series are standardized before they are used for clustering. For standardization, we rely only on the statistics obtained from the training set, keeping the global statistics completely hidden from the algorithms in the training process. This practice, referred to as \textit{local normalization} \cite{ahmadzadeh2021how} makes the reported performance as realistic as possible, since in operational settings, global statistics are unknown statistics.

    \subsection{Flare-Forecast Evaluation Methods}\label{subsec:evaluation}
        Typical forecast metrics used for deterministic performance verification in flare forecasting models are the True Skill Statistic (TSS) \cite{hanssen1965relationship} and a realization of the Heidke skill score (HSS) \cite{balch2008updated} based on its original definition introduced in \cite{heidke1926berechnung}. Let $tp$ denote the count of true-positives (correctly classified flaring instances), $tn$ denote the count of true-negatives (correctly classified non-flaring instances), and similarly, let $fp$ and $fn$ denote false-positive counts (misclassified as non-flaring) and false-negatives counts (misclassified as flaring). Then, TSS is defined as the difference between the probability of detection, $\tfrac{tp}{p}$, and the probability of false alarm, $\tfrac{fp}{n}$. Equivalently, $\textrm{TSS}=\frac{tp}{p} - \frac{fp}{n}$, where $p = tp + fn$ and $n = fp+tn$ are the numbers of the positive and negative instances, respectively. The value of TSS lies in the range $[-1,1]$: a score of $-1$ indicates that the model's predictions are entirely incorrect, $0$ reflects no predictive skill, and $+1$ represents a perfect model that correctly classifies all instances. HSS (referred to as HSS2 in some papers \cite{ahmadzadeh2021how, bobra2015solarflare}) is the other metric which quantifies the performance of a model by comparing it to the random-guess model. This is formulated as $\frac{2 ((tp \cdot tn) - (fn \cdot fp))}{p (fn + tn) + n (tp + fp)}$. HSS2 is interpreted same as TSS with higher values indicating a better (than random) performance.

        For balanced data, TSS equals HSS. For imbalanced data, HSS penalizes misclassification of the minority class (flaring instances) more than that of the majority class. So, the key difference between TSS and HSS is that unlike TSS, HSS is a function of class-imbalance ratio. This is critical because when comparing models tested on datasets with different imbalance ratios, only using TSS would be meaningful, however, the absence of HSS obscures the true performance of models. Generally, a simultaneous increase in TSS and HSS is desired, instead of observing an increase in TSS at the cost of HSS. This is the reason that it is strongly recommended that these metrics are used in pairs \cite{Bloomfield_2012, bobra2015solarflare} (also see the detailed discussion in \cite{Leka2019comparison}). This skewed performance is the reported trend in the literature when models perform suboptimally.

        Rand Index (RI) and Adjusted Rand Index (ARI) are common external measures of cluster quality. External measures evaluate cluster quality by comparing known labels ($y$) to cluster labels ($\hat{y}$). RI measures cluster quality based on the similarity of $y$ and $\hat{y}$. ARI does the same, but is adjusted for random chance. We select ARI over RI because of its interpretability and its invariance to $k$ value \cite{arinik2021characterizing}. ARI scores range from [-1,1], where +1 indicates perfect cluster assignment, 0 indicates a random guess model and -1 indicates worse than random cluster assignments.

\section{Experiments Towards Most Optimal}\label{sec:experiments}
    To study the effectiveness of the distance measures on classifying flare time series (using clustering algorithms), we first find the optimal configuration. This concerns the settings of the clustering algorithms as well as the parameters defining each distance measure. This optimal configuration---even if it does not yield a competitive performance compared with the best models reported in the literature---makes it possible to investigate the effectiveness of the high-dimensional distance measures. Therefore, in these experiments, while we optimize the performance of $k$-medoids algorithm (including the distance metrics used), we consider all other factors as extraneous variables. For example, sampling of instances, augmentation of data, or preprocessing of time series could each potentially enhance the performance, however, such steps obscure examining the direct contribution of distance measures used.

    We chose clustering, specifically because they directly rely on the effectiveness of the distance measures. Using clustering algorithms other than $k$-medoids certainly have merits, but the hypothetical gain in performance would then be attributed to the clustering strategy not the effectiveness of the distance measures. Therefore, we do not expand our experiments beyond one clustering algorithm.

    In short, in this section, we run 4,099 experiments to find the optimal configuration corresponding to the highest flare-forecast performance on SWAN-SF. This configuration pertains to the $k$-medoids clustering algorithm and the three distance measures, namely DTW, MSM, and TWE. A repository containing the experiments described in this paper is available at this link\footnote{\url{https://bitbucket.org/dataresearchlab/clusteringflares_icdmw25/src/main/}}.

    \begin{figure*}
        \centering
        \includegraphics[width=1\linewidth]{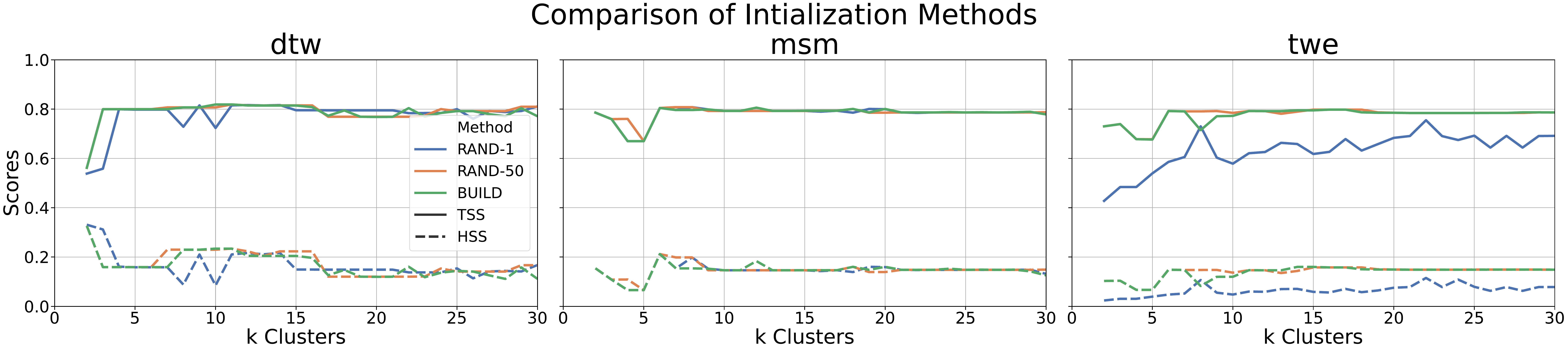}
        \caption{\footnotesize Partition 2 validation results for $k$-medoids. Each plot shows TSS and HSS scores for three initialization methods and one distance measure.}
        \label{fig:initVal}
    \end{figure*}
    
    \subsection{Experimental Settings}\label{subsec:settings}
        We optimize the clustering model separately for each distance measure. For all experiments we use a simple $k$-medoids model, using the ``Faster Pam'' algorithm \cite{schubert2021fastandeager} and distances are computed from a single pre-processed data set (see Sec.~\ref{sec:data_and_evaluation_methods}). P1 is selected for training, P2 for parameter tuning (validation set), and P3 for evaluating (test set) the performance of the trained model. The models are evaluated against the ground-truth labels, specifically against the flaring class FL and non-flaring class NF. We score models with TSS, HSS, and ARI. Optimal settings are defined as those settings which maximize both TSS and HSS. Pairwise distances are computed independently, as described in Sec.~\ref{sec:distances} and with the default parameters implemented in the AEON package \cite{aeon2024middlehurst}, except where otherwise noted. The default parameters are as follows: For DTW $w=1.0$, for MSM $w=1.0$, $c=1.0$, and for TWE $w=1.0$, $\nu=0.001, \lambda=1.0$.

        To evaluate cluster labels against ground-truth labels, we must first map them to their most likely class label. The class map is created during the training phase and then reused for the prediction phase. This ensures predictions are evaluated based on the train partition and not influenced by the validation or test partitions. Each medoid is mapped to the class that minimizes within-cluster $fp$ instances (equivalent to maximizing within-cluster $tp$ instances). To account for imbalance within the subclasses, the mapping is created with a normalized contingency matrix. We do not restrict $k=n$ because we do not expect the underlying clusters to have the spherical shape easily captured with a basic clustering algorithm. Due to the complexity of the data set as well as the underlying subclasses, we consider the possibility that allowing multiple medoids to represent a single class will capture a larger portion of the true clusters. This mapping strategy allows for a one to many mapping, ensuring that every cluster is mapped to one class. Once mapped, clusters are scored with TSS, HSS, and ARI.

    \subsection{Evaluation of Initialization Methods}

        Initial medoid selection can impact the quality of the final clusters. Good medoid initialization selects data points so that they are distributed near to each of the true clusters. In general, the nearer the initial medoids are to the true cluster centers, the more likely we are to find the optimal clustering. A good initialization strategy can also reduce the overall clustering time. For our experiments, we use the open-source Fast $k$-medoids package \cite{schubert2022fastrust}. This package offers three initialization strategies, namely, random, build, and first. First simply selects the first $k$ data points for its medoids. In general, this strategy results in poor initialization as it does not ensure that the medoids are spread over the dataset. Random is similarly simplistic, but in this case $k$ medoids are selected at random from the entire dataset. The stochastic nature of the approach makes it more likely that initial medoids will be distributed over some or all of the true clusters. Random medoids are more likely to be selected from dense areas, simply because there are a higher percentage of points in that region. \cite{holder2023clustering}. It is common practice to implement random initialization with 10 restarts and use the medoid initialization that minimizes the error. Build is the first step in the original PAM algorithm \cite{kaufman1990pam}. Build is a greedy approach which begins by selecting the medoid that minimizes the sum of distances for all data points and then choosing the next medoid that minimizes the previous sum. 
   
        This greedy approach is more computationally demanding, but can reduce overall clustering error.

        For this experiment, we consider only build and random methods. We test random with no restarts (RD-1), random with 50 restarts (RD-50), and build with no restarts (BUILD). The best medoids were selected based on which gave the minimum inertia. Initialization methods are then evaluated based on computation time as well as on TSS and HSS scores.

        \begin{table}[t]
            \centering
            \caption{\footnotesize
            Scores from Partition 3 after selecting an optimum $k$ from Partition 2. Methods that produced identical results are grouped into a single row for conciseness. The best result for each distance is shown in \textbf{bold}.}
            \label{tab:initTest}
            \begin{tabular}{r r l l l l}
                \toprule
                \textbf{Distance} & \textbf{Initialization Method} & \textbf{$\boldsymbol{k}$} & \textbf{TSS} & \textbf{HSS} & \textbf{ARI} \\ 
                \midrule
                DTW    & RD-1                             & 2  & 0.544          & 0.324               & 0.298           \\ 
                       & \textbf{RD-50, BUILD}            & \textbf{2}          & \textbf{0.577}      & \textbf{0.326}   & \textbf{0.298}  \\ 
                \addlinespace
                MSM    & \textbf{RD-1, RD-50, BUILD}    & \textbf{6}          & \textbf{0.729}      & \textbf{0.253}   & \textbf{0.210}  \\ 
                \addlinespace
                TWE    & RD-1                             & 22 & 0.683          & 0.140               & 0.084           \\ 
                       & \textbf{RD-50}                   & \textbf{15}         & \textbf{0.757}      & \textbf{0.208}   & \textbf{0.157}  \\ 
                       & BUILD                              & 6  & 0.737          & 0.207               & 0.158         \\ 
                \bottomrule
            \end{tabular}
        \end{table}

        Fig.~\ref{fig:initVal} shows TSS and HSS scores obtained for each initialization method while tuning on P2. 
        Interestingly, for MSM and DTW, Fig~\ref{fig:initVal} shows all three initialization methods produce identical or near identical scores throughout the tuning process. TWE is not as lucky, although RD-50 and BUILD overlap, RD-1 produces consistently and significantly lower scores. It is also worth noting that in some cases RD-50 took over 30 restarts to find the same optimal medoids that BUILD selected with no restarts. Evaluating for optimal $k$ on P3 produced the scores shown in Table~\ref{tab:initTest}. After examining the tuning results, it is not too surprising to see overlap in the scores for each initialization method. In the case of MSM, initialization method appears to have no impact on final score. RD-50 and BUILD produce the highest scores for DTW and MSM. For TWE, RD-50 marginally outperforms BUILD (+0.02 TSS, +0.001 HSS, +0.001 ARI). Considering that RD-1 significantly and negatively impacted TWE outcomes, we discard that method. Considering the remaining methods, we note that BUILD is significantly faster and its outcomes are nearly identical to RD-50. Based on these findings, we initialize the remaining experiments with BUILD.

        Now that we found the best initialization method, we search for the optimal number of clusters.

    \subsection{Finding the Optimal Number of Clusters}\label{subsec:optK}

        For this experiment we tune the clustering algorithm to obtain an optimal $k$ while distance parameters are left at their default settings. We use the experimental settings and evaluation strategy as described in Sec.~\ref{subsec:settings} and each model is trained for $k=\{2,3,\dots,100\}$. For each distance measure, the $k$ value that maximizes TSS and HSS after tuning on P2 is selected for evaluation on P3.

            \begin{figure}[t]
                \centering
                \includegraphics[width=1\linewidth]{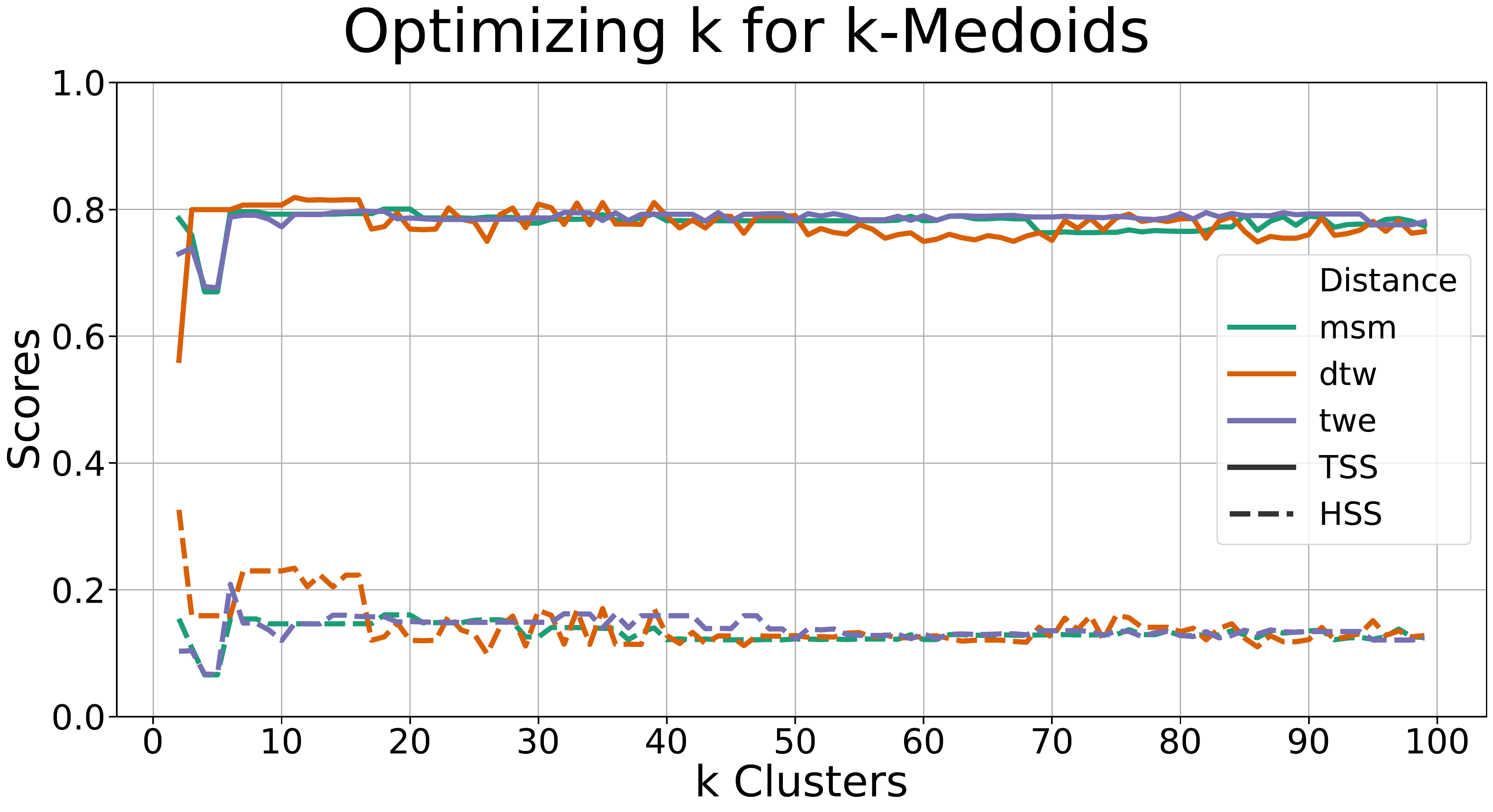}
                \caption{\footnotesize TSS and HSS scores for Partition 2 (validation partition) for $k = \{2,3,\dots,100\}$}
                \label{fig:optKtss}
            \end{figure}
    
            Fig.~\ref{fig:optKtss} shows TSS (solid) and HSS (dotted) for tuning on P2. All three measures perform similarly throughout optimization. DTW achieves the highest HSS scores, at $k=2$ and $k=\{8,9, \dots, 15\}$, but none of these scores surpass 0.35. Fig.~\ref{fig:optKtss} also shows that although $k=2$ produces DTW's maximum HSS, TSS is at its minimum. Maximizing HSS at the expense of TSS is not desirable. The best $k$ for each measure is $k=11$ for DTW and $k=6$ for MSM and TWE. In general, we note the best HSS gains for all measures occur when $k \le 15$.
            We use the findings from this experiment to narrow the search space for $k$ in the subsequent experiment, optimizing each distance (Sec.~\ref{subsec:optDist}).

    \subsection{Finding Optimal Parameters for Metrics}\label{subsec:optDist}
        \begin{table}[t]
        \caption{\footnotesize 
            Hyperparameter grid for each distance measure. Settings for $w$ are as follows. The Sakoe-Chiba band and Itakura Parallelogram are tested over the same range $w_{sc} = w_{ip} = \{0.02, 0.04, 0.05, 0.10, 0.15, 0.20, 0.30\}$. $w_{g} = \{1.0\}$ is the case of unconstrained (global) warping.
        }
        \centering
        \begin{tabular}{r l l l}
            \toprule
                                & \multicolumn{3}{c}{\textbf{Hyperparameter Grid for $k$-medoids}} \\ 
                                \cmidrule(lr){2-4}
            \textbf{Distance}   & \textbf{Clusters ($\boldsymbol{k}$)}   & \textbf{Window ($\boldsymbol{w}$)}        & \textbf{Additional Hyperparameters} \\
            \midrule
            EuD                 & $\{2,3,\dots,15\}$   & ---                               & --- \\
            \addlinespace
            DTW                 & $\{2,3,\dots,15\}$   & $w_{g},w_{sc}, w_{ip}$            & --- \\
            \addlinespace
            MSM                 & $\{2,3,\dots,15\}$   & $w_{g}, w_{sc}, w_{ip}$           & $c = \{10^{-2}, 10^{-1}, \dots, 10^{2}\}$ \\
            \addlinespace
            TWE                 & $\{2,3,\dots,15\}$   & $w_{g}, w_{sc}, w_{ip}$           &  \begin{tabular}[t]{@{}l@{}}
                                                                                                   $\nu = \{10^{-5}, 10^{-4}, \dots, 10^{0}\}$ \\
                                                                                                   $\lambda = \{0, 0.25,0.50,0.75,1\}$
                                                                                               \end{tabular} \\

            \bottomrule
        \end{tabular}
        \label{tab:paramGrids}
        \end{table}

        \begin{table}[t]
            \centering
            \caption{\footnotesize
                Comparison of distance measures with their optimal hyperparameters and performance scores.
            }
            \label{tab:optDist}
            \begin{tabular}{r l l l l l}
                \toprule
                                    & \multicolumn{5}{c}{\textbf{Optimal Results for $k$-medoids}} \\ 
                                        \cmidrule(lr){2-6}
                \textbf{Distance}   & $\boldsymbol{k}$      & \textbf{Hyperparameters}      & \textbf{TSS}  & \textbf{HSS}  & \textbf{ARI} \\ 
                \midrule
                EuD                 & 11     & ---                               & 0.778     & 0.275      & 0.167   \\ 
                \addlinespace
                DTW                 & 11     & $w_{sc}=0.15$                     & 0.766     & 0.263      & 0.218   \\ 
                \addlinespace
                MSM                 & 5      & $w_g$, $c=10.0$                   & 0.754     & 0.268      & 0.224   \\ 
                \addlinespace
                TWE                 & 6      & $w_g$, $\nu=1.0$, $\lambda=1.0$   & 0.729     & 0.253      & 0.210   \\ 
                \addlinespace       
                MDD                 & 6      & $\epsilon=\{1,2\}$                & 0.685     & 0.280      & 0.242   \\ 
                \bottomrule
            \end{tabular}
        \end{table}
    
        For this experiment we focus on tuning the hyperparameters of each distance measure to obtain three optimized models. 

        This time we train the models for a range of hyperparameters and $k$ values, to cover all combinations. Ranges for hyperparameters are selected based on recommendations in the literature. For MSM, the authors note that tuning with a few widely spaced values was sufficient to produce a model with competitive results \cite{stefan2013msm}. Based on the findings from the previous experiment in Sec.~\ref{subsec:optK}, we consider $k\in\{2,3,\dots,15\}$ in the grid search. After an exhaustive search of hyperparameter combinations, the $k$ value and distance settings that maximize HSS (and equivalently, give a balance between TSS and HSS) are selected, based on validation partition results. In case of ties, we follow standard practices where possible. For DTW we break ties by choosing the lower $w$, for TWE we first reduce ties by maximizing $\nu$, if any ties remain we then maximize $\lambda$ \cite{marteau2009twe}.  
        We report the results obtained from the testing phase. The final clusters are scored based on TSS, HSS, and ARI.

        In addition to these hyperparameters, the AEON package implements an optional ``window'' parameter for MSM and TWE, with the same functionality as DTW's ``window'' parameter. We include this parameter in our search space. 
        We include the special case when $w=0$ for DTW (equivalent to EuD), meaning EuD performance is considered during optimization. Additionally, because MSM and TWE do not have explicit window size recommendations, we use the recommended $w$ settings given for DTW and include $w=1$ to include performances that reflect their original implementations. See Table~\ref{tab:optDist} for the optimized settings and scores.


    \begin{table*}
            \caption{\footnotesize Changes in performance (percent-correct) of $k$-medoids algorithms from training on Partition 1 to testing on Partition 3.}
            \centering
            \begin{tabular}{rccccccccccc}
                \toprule
                & \multicolumn{11}{c}{Per-cluster Analysis for Optimal $k$-medoids with \textbf{DTW}} \\ 
                \cmidrule(lr){2-12}
                Actual Label & X & M & FQ & X & M & X & M & M & X & C & FQ \\
                Assigned Label & FL & FL & NF & FL & FL & FL & FL & FL & FL & NF & FL \\
                \%-correct (train) & 96.14\% & 98.28\% & 99.47\% & 95.45\% & 86.75\% & 100.00\% & 100.00\% & 98.80\% & 87.64\% & 54.67\% & 71.81\% \\
                \%-correct (test) & 30.44\% & 55.73\% & 99.98\% & 53.70\% & 11.89\% & 54.11\% & 30.62\% & 25.52\% & 17.28\% & 98.40\% & 9.12\% \\
                \bottomrule
                \addlinespace
            \end{tabular}
            \begin{tabular}{rccccccccccc}
            \toprule
                                    & \multicolumn{5}{c}{Per-cluster Analysis for Optimal $k$-medoids with \textbf{MSM}}    & \multicolumn{6}{c}{Per-cluster Analysis for Optimal $k$-medoids with \textbf{TWE}}\\
                \cmidrule(lr){2-6}\cmidrule(lr){7-12}
                Actual Label        & C         & M         & FQ        & M         & M                  & M         & FQ        & M         & FQ        & C         & M \\
                Assigned Label      & NF        & FL        & NF        & FL        & FL                 & FL        & FL        & FL        & NF        & NF        & FL \\
                \%-correct (train)  & 55.80\%   & 98.24\%   & 99.14\%   & 94.31\%   & 84.16\%            & 100.00\%  & 82.15\%   & 89.51\%   & 99.15\%   & 50.72\%   & 97.54\% \\                
                \%-correct (test)   & 98.07\%   & 34.54\%   & 100.00\%  & 26.89\%   & 13.31\%            & 46.82\%   & 10.77\%   & 17.27\%   & 100.00\%  & 97.67\%   & 33.24\% \\ \hline                
            \end{tabular}
            \label{tab:percent_correct_all_models}
        \end{table*}
        
\section{The Root Causes of Suboptimal Performance}
    As our experiments showed, the most optimal $k$-medoids models could not surpass the stagnant performance reported in many papers in the past decade or so. Considering a balance between TSS and HSS, the best model reached $\text{TSS}=0.754$ and $\text{HSS}=0.268$---unimpressive! That said, we can now examine one root cause of this issue, i.e., the ineffectiveness of high-dimensional metrics on stochasticity of flare time series.

    \subsection{Analysis of Clusters}
        The per-cluster quantities are listed in Table~\ref{tab:percent_correct_all_models}. There is a very clear pattern in how the percent-correct numbers change. Recall that ``true label'' of each cluster is assigned based on its medoid's actual label. The ``assigned label'' is given by the class label of the majority of instances in that cluster, during training. Looking at the numbers, there is a very clear pattern present: when the assigned label of a cluster is NF (non-flaring), on test set, the cluster either maintains or gains power, i.e., percent-correct improves. When the assigned label of a cluster is FL (flaring), the opposite trend occurs; its power drops significantly on test set. This is because the test set is extremely imbalanced (unlike the undersampled training set) and therefore, a disproportionate number of NF instances are added to all clusters. This saturates the less-populated FL clusters, reducing their power, while boosting that of the NF clusters. This should not come as a surprise, as the extreme class-imbalance issue impacts all (supervised and unsupervised) algorithms, although the technical reasoning might be different.

    \subsection{Improvement over EuD}
        As mentioned in Sec.~\ref{subsec:optDist}, among many experiments, we also ran $k$-medoids with DTW where $w=0$. This is equivalent of using EuD. This achieved $\text{TSS}=0.778$ and $\text{HSS}=0.275$, which is just as good as the best models we found using DTW, MSM, and TWE. This conclusively indicates that none of those advanced distance metrics could find similarity patterns beyond the little that could be found by EuD's simple, non-elastic approach. This could be interpreted in two ways---both fall outside the scope of our paper, but we mention for context. One would wonder that there may be a different approach to capture the similarity between high-dimensional data points of flares. One would also wonder if flare time series exhibit any quantifiable similarity patterns to yield performance beyond the current state of flare forecast. Each of these ideas merit further research. Note that we do not claim our $k$-medoids' performance competes with the best ones reported in the literature. As mentioned earlier, this combination of TSS and HSS reveals the primitive nature of this model. However, the fact that despite all optimization efforts, none of the elastic metrics could outperform EuD, indicate that there is little similarity that can be captured by DTW, MSM, and TWE, when we deal with SWAN-SF and the stochasticity it entails.

        \begin{figure*}
            \begin{subfigure}[h]{0.5\linewidth}
                \includegraphics[width=\linewidth]{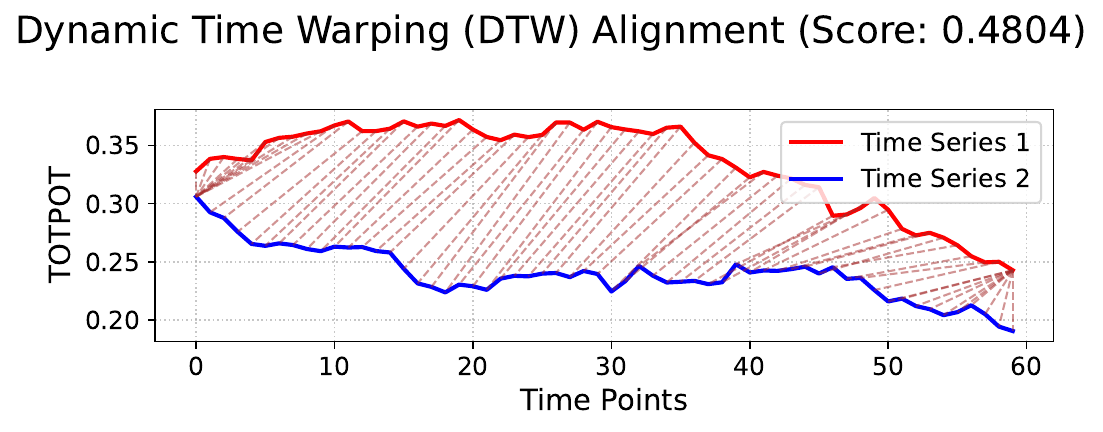}
                \label{fig:dtwalignP}
            \end{subfigure}
            \hfill
            \begin{subfigure}[h]{0.5\linewidth}
                \includegraphics[width=\linewidth]{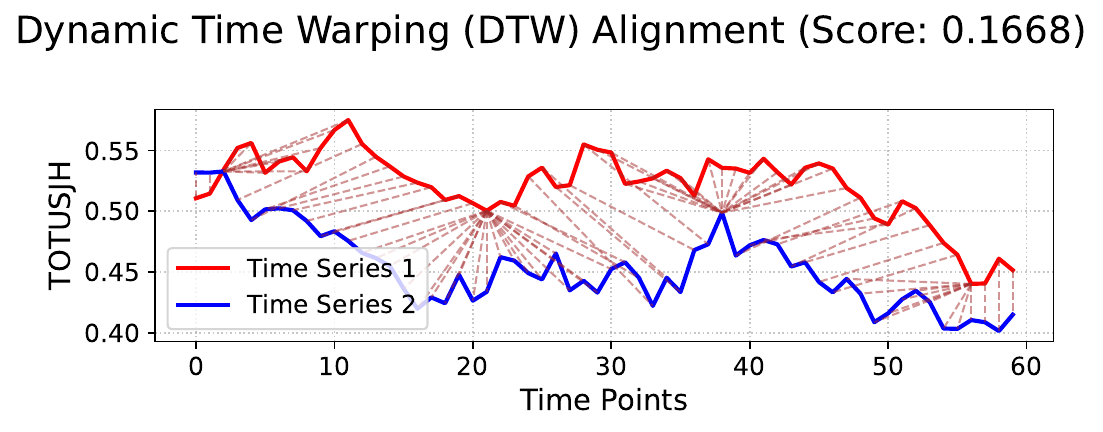}
                \label{fig:dtwalignH}
            \end{subfigure}
            \caption{\footnotesize Visualization of point mapping for DTW applied on two times series of SWAN-SF with label FL, parameters are TOTPOT and TOTUSJH.}
            \label{fig:qual_anal}
        \end{figure*}

        While the similar performance of EuD to that of DTW, MSM, and TWE shows statistically that the other (more advanced) distance metrics do not contribute more than EuD, we looked at the exact point-mapping of those metrics on examples throughout the test set. In our qualitative analysis two main scenarios stand out to us: (1) the distance metric often performs non-elastic mapping, i.e., what EuD is designed to do, (2) DTW exhibits pathological warping. Examples of these cases are shown in Fig.~\ref{fig:qual_anal}. When non-elastic mapping is used, we conclude that the elastic metrics either could not find patterns or the correct pattern did not require elastic mapping. Regardless of the correctness of label assignment in clustering, EuD could have achieved the same. When elastic mapping is used but the warping is pathological, it is likely that EuD would have had the same chance (if not higher) of assigning the correct label. Although we empirically observed these two patterns very frequently, in the absence of an algorithm to count the number of warpings and the number of pathological warpings, we cannot statistically verify this claim.


\section{Conclusion and Future Work}
    This study evaluated the performance of three prominent elastic distance measures for time series clustering on the SWAN-SF dataset. Our experimental results demonstrate that these measures exhibit limited efficacy in surpassing the performance of the standard Euclidean distance baseline. Despite the application of extensive parameter tuning strategies, only marginal performance enhancements were observed. This finding suggests that further optimization on these measures is likely to yield diminishing returns. This justifies research on the design of more novel high-dimensional distance measures, particularly those which do not rely on the point matching strategy, such as  \cite{ahmadzadeh2022ts, khazaei2024multiscale}.

\section*{Acknowledgment}
    This material is based upon work supported by the National Science Foundation under Grant No. 2209912 and 2433781, directorate for Computer and Information Science and Engineering (CSE), and Office of Advanced Cyberinfrastructure (OAC), and Grant No. 2511630, AST Division Of Astronomical Sciences and MPS Directorate for Mathematical and Physical Sciences.

\bibliographystyle{splncs04}
\raggedright
\vspace{-0.3cm}
\bibliography{main}

\end{document}